\documentclass[a4paper]{scrartcl}
\usepackage{amssymb}
\usepackage{amsmath}
\usepackage{amsthm}
\usepackage{latexsym}
\usepackage{enumerate}
\usepackage{graphicx,color}
\usepackage{dsfont}
\usepackage{float}
\usepackage[textwidth=17mm]{todonotes}
\usepackage{placeins}
\usepackage[title]{appendix}
\usepackage{etoolbox}
\patchcmd{\appendices}{\quad}{: }{}{}
\usepackage[normalem]{ulem}
\usepackage{lmodern}
\usepackage{geometry}
\usepackage{enumitem}
\usepackage{xcolor}
\usepackage{framed}
\usepackage{hyperref}
\usepackage{authblk}

\usepackage{tikz}
\usetikzlibrary{arrows,positioning,calc} 
\tikzset{
    >=stealth',
    punkt/.style={
           rectangle,
           rounded corners,
           draw=black, thick,
           text width=5em,
           minimum height=2em,
           text centered},
    punkts/.style={
           rectangle,
           rounded corners,
           draw=black, thick,
           text width=3em,
           minimum height=2em,
           text centered}, 
    punktl/.style={
           rectangle,
           rounded corners,
           draw=black, thick,
           text width=7em,
           minimum height=2em,
           text centered},
    pil/.style={
           ->,
           shorten <=4pt,
           shorten >=4pt,},
    pildotted/.style={
           ->,
           shorten <=4pt,
           shorten >=4pt,
  dotted,},
      pildashed/.style={
           ->,
           shorten <=4pt,
           shorten >=4pt,
  dashed,
  }
}

\newcommand*\diff{\mathop{}\!\mathrm{d}}
\newcommand{\indep}{\perp \!\!\! \perp}

\usepackage[round,authoryear]{natbib}
\newcommand*{\doi}[1]{\href{http://dx.doi.org/#1}{doi: #1}}
\usepackage{mdframed}
\usepackage{multirow}
\usepackage{array}
\usepackage[font=small,labelfont=bf]{caption}

\title{Loss of earning capacity in Denmark -- an actuarial perspective} 

\author[1,$\star$]{Christian Furrer}
\author[1,2]{Oliver Lunding Sandqvist}

\affil[1]{\footnotesize Department of Mathematical Sciences, University of Copenhagen, Universitetsparken 5, DK-2100 Copenhagen \O, Denmark.}
\affil[2]{\footnotesize PFA Pension, Sundkrogsgade 4, DK-2100 Copenhagen \O, Denmark.}
\affil[$\star$]{\footnotesize Corresponding author. E-mail: \href{mailto:furrer@math.ku.dk}{furrer@math.ku.dk}.}

\date{\vspace{-8mm}}

\begin{document}

\maketitle

\begin{abstract}

We describe challenges and opportunities related to risk assessment and mitigation for loss of earning capacity insurance with a special focus on Denmark. The presence of public benefits, claim settlement processes, and prevention initiatives introduces significant intricacy to the risk landscape. Accommodating this requires the development of innovative approaches from researchers and practitioners alike. Actuaries are uniquely positioned to lead the way, leveraging their domain knowledge and mathematical-statistical expertise to develop equitable, data-driven solutions that mitigate risk and enhance societal well-being.

\end{abstract}

\vspace{5mm}


\section{Introduction} \label{sec:introduction}

The purpose of insurance is to protect against uncertain events that lead to financial losses, which is first and foremost achieved by the pooling of diversifiable risks. Disability benefits play a crucial role in ensuring income stability for individuals by reducing financial vulnerability during periods with reduced earning capacity as well as supporting part-time employment. The latter relates to the fundamental right to work. This right may be found in the Declaration of Philadelphia (1948)\footnote{Declaration concerning the aims and purposes of the International Labour Organisation.}, in Article~23 of the Universal Declaration of Human Rights (1948), in Article~10 of the Valencia Declaration (1998)\footnote{Declaration of Human Duties and Responsibilities.}, and in Article~15 of the EU Charter of Fundamental Rights (2000). In a welfare state, the right to work is matched by a duty to work for those that are able -- and support for the disabled and disadvantaged. Such is the situation in Denmark, where a well-rounded three-pillar system consisting of the public sector, labour market pensions, and private insurers offers a rich interplay between public benefits and insurance coverage \citep[][Chapter 6]{Ekspertgruppe:2022}. The insurance coverages play a vital role in today's society, with approximately one in five Danes expected to rely on disability insurance at some point during their careers\footnote{\url{https://danicapension.dk/en/personal/your-insurances/all-insurance-options/loss-of-earning-capacity} (accessed 14/1/2025).}.

For classic risks such as mortality and longevity, a wide range of proven modeling approaches are available to the actuary. In comparison, disability insurance presents unique challenges. Disabilities can v{a}ry in form, degree, and duration, and in the interplay with the public system, leading to complications in risk assessment and management.  There is untapped potential in this field -- and this potential can be harnessed by actuaries due to their unique combination of domain knowledge and mathematical as well as statistical competence.

We attribute the relatively unrealized potential to historical reasons first and foremost. Technological progress has just recently allowed the storage and processing of much larger quantities of data than in the past, which combined with advances in data analytics, machine learning, and task automation has greatly expanded the actuaries' and statisticians' sphere of influence. This also leads to increased competition in the market, both in relation to product design, pricing, and value creation, which puts additional pressure on actuaries to fulfill the potential.

To keep the presentation concise and closely aligned with our practical experience, we focus on the Nordic countries, especially Denmark. However, we expect the discussions to also be relevant for actuaries in other regions. For a broad introduction to the German health insurance landscape and actuarial practices, confer with~\citet{Milbrodt:Rohrs:2016}. 

{
The contribution of the paper is two-fold. First, we perform a comparative review of the existing literature, highlighting key insights for actuaries that presently resides in more mathematically demanding texts. Second, we analyze aspects of disability insurance that are essential for accurate risk modeling, but which so far have received limited attention in the literature; and in connection to this, we highlight directions for future research.}

The paper is structured as follows. In Section~\ref{sec:risks}, we describe and characterize the risk environment for loss of earning capacity. Subsequently, Section~\ref{sec:Product} contains a formalization and comparison of existing insurance coverage designs with a focus on the Danish disability insurance market. Based hereon, the next two sections are devoted to the topics of risk assessment through prediction and risk mitigation through prevention, respectively. In Section~\ref{sec:Modeling}, the adequacy of current approaches to stochastic modeling, pricing, and reserving of disability risk is assessed. Next, in Section~\ref{sec:prevention}, methods for impact evaluation of prevention initiatives are discussed. {Data availability is a key concern, and practical challenges associated with disability insurance data are discussed in Section~\ref{sec:data}.} Throughout, we identify avenues for further research and improved actuarial practice. Closing remarks are provided in  Section~\ref{sec:Conclusion}.

\section{Risks and agents} \label{sec:risks}

This section describes the different risks and agents involved in disability insurance, highlighting the complex interplays that arise. The main purpose is to identify and provide an overview of operational areas for which actuarial expertise is crucial to ensure successful risk assessment and mitigation. For this reason, we mainly place ourselves in the position of the insurer and the actuary. However, some of the insights we provide could also be relevant in a broader context, for instance in the design and implementation of public welfare reforms.

\subsection{Background}\label{subsec:background}

Denmark is, like its northern neighbors, a welfare state with an elaborate social safety net, universal healthcare, and collective bargaining~\citep{PedersenKuhnle:2017}. The social safety net seeks to cultivate the economic and social well-being of citizens with inadequate or no income, not least due to temporary or permanent incapacity to work. The wide range of public benefit schemes reflects efforts to balance measures that facilitate strengthened workforce participation with citizen welfare. This is not to insinuate that these two objectives are necessarily opposed to each other; on the contrary, there is a rich interplay between them. According to~\citet{BratsbergFevangRoed:2013}, for example, in Norway job loss is a major cause of disability program entry. In Denmark, public benefit schemes include temporary sickness benefits, but also long-term benefits like so-called flexjob benefits as well as early retirement pension. The {Danish} system is widely recognized to be complex, frequently attracts political attention, and is under continuous development~\citep{Ekspertgruppe:2024,Beskæftigelsesministeriet:2025}. {The situation in Sweden is comparable, with~\cite{Djehiche:Lofdahl:2018} highlighting the 2008 social security reforms and the later rollback pertaining to the national sickness insurance system.}

Despite the extensive social safety net that characterizes the Nordic model, some individuals who become partly or completely incapacitated to work experience a considerable loss of income and inability to sustain their pension contributions. This is especially the case for high earners, but is expected to become increasingly relevant to the wider population. In Denmark, the increased privatization of welfare benefits can be seen as part of a larger movement towards a perhaps fundamentally different society characterized by its rich interplay between private and public actors and mixed solutions. We refer to~\citet{FischerKvist:2023} for a recent in-depth review and discussion of this trend.

In many ways, the key role of insurance is to ensure stability of an individual's income and consumption by risk-sharing through time and across individuals~\citep{Pensionspanelet:2019}. In light of the aforementioned societal shifts, it is then no surprise that the market for so-called loss of earning capacity insurance (or for the sake of linguistic simplicity: \textit{disability insurance}) has been growing and continues to grow. In Denmark, this includes private individual schemes managed by commercial insurance companies and supplements to company or occupational pension schemes offered by commercial or cooperative pension providers. Here is also the statutory workers' compensation insurance partly managed by the so-called Labour Market Insurance (Arbejdsmarkedets Erhvervssikring, AES) in accordance with the Workers' Compensation Act. The fact that the insurance schemes aim to ensure consumption smoothing entails an unfortunate spillover effect of complexity. Given that the public benefits are {both} supposed to and {indeed} constitute the main pillar, the design and appropriateness of insurance coverages must be considered residual hereto and under penalization by offset rules as to not under- or overcompensate the insured, see also Sections~3.1 and~5.2 in~\citet{Pensionspanelet:2019}. This has given rise to complicated insurance coverages that interact closely with the public benefits, inheriting not least the complexities of the latter. For an overview of the Danish disability insurance market, see also Chapter~6 in~\citet{Ekspertgruppe:2022}.

Disability risk and insurance are also just from a purely biometric perspective more intricate than, say, mortality risk and life insurance. They are both long-term risks, but death is, after all, categorical, while disability comes in various forms and degrees. Proper risk assessment therefore necessitates an effective collaboration between case managers and medical professionals. Furthermore, while mortality risk has developed in a rather stable way, at least disregarding pandemics, disability risk has historically developed in a more volatile manner that directly affects disability insurers. An example hereof is the major increase in mental health claims, with the Danish insurer Velliv in 2022 reporting that 70 \% of its payments to young disability claimants were stress related\footnote{\url{https://finanswatch.dk/Finansnyt/Pension/Velliv/article13748200.ece} (accessed 14/1/2025).}.

All in all, this paints a rather complicated picture of disability insurance in Denmark and the other Nordic countries. To exemplify and systematize the complexities, we introduce and study two realistic cases based on our actuarial experience in the Danish insurance and pensions industry. The cases are intended to reflect reality, while also bringing to light characteristics about the product and its risks that are of particular significance from an actuarial perspective.

\subsection{Alex and Charlie}\label{subsec:cases}

In the following, we present two fictitious, but quite illustrative, disability cases. The first case, about Alex, describes a possible trajectory in the public system and underscores the raison d'être for disability insurance in Denmark. The second case, about Charlie, focuses on the type of information that becomes available during the insurer's claims processing. The idea is to provide a better understanding of the data that the actuary can actually access for dynamic decision making and for actuarial and statistical modeling purposes.

\begin{mdframed}

\textbf{Alex's accident}

Alex, 33 years, falls on their bike on the way home from work and hits their head. In the first few weeks after the accident, they experience frequent headaches, but their work life is not immediately affected. In the next two months this changes as their condition deteriorates. They go on sick leave for a couple of months during which they receive full pay from their employer; the employer is partially compensated by the municipality. Alex's family physician assesses that their earning capacity has been reduced to one-third in their current job.${}^{\textrm{a}}$

Alex attempts to continue to work at their current employer, but only for 12 hours a week. However, this scheme is unsuccessful and they end up being terminated. For a while, Alex's only source of income is therefore sickness benefits combined with savings. The sickness benefits are eventually discontinued, at which time the municipality enrolls Alex in a vocational assessment and resource clarification program. During this program, they receive so-called resource benefits from the municipality.${}^{\textrm{b}}$

After about two years, Alex and the municipality identify a more appropriate career choice for Alex. The new job pays less than their original job, but they are able to work half of normal hours for full pay and they receive a so-called flex job salary supplement from the municipality.${}^{\textrm{c}}$

Alex worries about the rising inflation and the size of their pension since they had several years where they withheld pension contributions and are now paying a lower amount than before. Alex still suffers from severe and crippling headaches twice or thrice every month.${}^{\textrm{d}}$
\end{mdframed}

${}^{\textrm{a}}$If Alex had had disability insurance coverage, they could have applied for disability benefits supported by their medical report. If awarded, their employer would be additionally compensated by insurance benefits, however in such a way that the total compensation does not exceed Alex's original salary. This means that it would be less expensive for the employer to retain Alex and, in the longer term, uncover alternative work arrangements.

${}^{\textrm{b}}$Resource benefits are generally lower than sickness benefits, so Alex's financial situation worsens as time passes, further negatively impacting their quality of life. The public benefits could have been complemented by insurance benefits, would Alex have been covered.

${}^{\textrm{c}}$Note that by this point, Alex has received three different types of public benefits, each subject to differing criteria: sickness benefits, resource benefits, and flex job salary supplement. Several other types of benefits exist, indicating the complexity of the public system. Even if Alex had been able to work normal hours, a disability insurance could have continued to play some role, given the difference in salary between the new and the original job.  This is because disability insurance benefits typically are based on the original salary just before disability.

${}^{\textrm{d}}$Premium exemption insurance, also known as premium waiver insurance, would ensure the continuation of pension contributions at the original level in case of a disability. This coverage aims to maintain the same level of welfare for the pensioner regardless of their disability history before retirement. Both the premium exemption insurance and the disability insurance may be subject to suitable indexation as to mitigate the influence of inflation.

\begin{mdframed}

\textbf{Charlie's claim}

The insurer receives a disability claim from Charlie, a 45 year old primary school teacher, on the ground of their reduced earning capacity. Charlie has recently returned from a three months sick leave due to work-related stress and burnout, and is currently only working part time, around 16 hours a week.${}^{\textrm{e}}$
 
The insurer’s claims processing team reviews Charlie’s medical records and rejects the claim, citing insufficient documentation of at least a 50 \% reduction in earning capacity, which is an eligibility requirement stipulated in the insurance contract. The rejection is communicated to Charlie about four weeks after they submitted their claim.${}^{\textrm{f}}$

Several months later, Charlie reapplies with new medical documentation from their family physician as well as an independent psychiatrist, detailing severe stress and early-stage depression, which has led Charlie to now only working 12 hours a week. The insurer reassesses the claim and approves benefits retroactively, starting three months after the first day of Charlie' initial sick leave to take into account the waiting period stipulated in the insurance contract.${}^{\textrm{g}}$

The insurer regularly reviews Charlie’s progress. After a doctor concludes that Charlie can return to work full hours, the insurer ceases benefits. However, a second medical opinion shows no improvement, and the insurer hence resumes payments.${}^{\textrm{h}}$

Over time, Charlie's condition worsens into a serious depression, and Charlie continues to receive insurance benefits. The insurer eventually refers Charlie to a vocational assessment. Charlie is reluctant to participate, citing a lack of energy. However, the insurer informs Charlie that the program is a prerequisite for receiving benefits, after which Charlie follows the recommendation.${}^{\textrm{i}}$

It is concluded that a job as a librarian at a nearby library would be optimal for Charlie's condition and Charlie fortunately manages to secure the job. The insurer regularly obtains updates from Charlie over the next 18 months as their mental health and work hours increase. Finally, Charlie reports that they feel much better and are now working full hours at the library, leading the insurer to cease all benefits.${}^{\textrm{j}}$

The next time the insurer hears from Charlie is at their retirement age. They are claiming their pension and look forward to spending more time with their grandchildren.

\end{mdframed}

${}^{\textrm{e}}$There has thus been a reporting delay of more than three months since the onset of illness. Long reporting delays are often observed for disabilities and for multiple reasons, including the illness in itself, the need to collect medical documentation, an inattentiveness or unawareness of the insurance coverage, and the lack of a short-term monetary incentive due to the availability of sick leave.

${}^{\textrm{f}}$This constitutes a rejection of the claim, however it turns out to only be a temporary rejection given that Charlie later successfully reapplies. Temporary rejections are common as it is difficult to precisely determine the degree of loss of earning capacity. The involvement of medical professionals is required, and the counterfactual nature of the question of how many hours the insured \textit{might be able to} work invokes an inevitable element of subjectivity.

${}^{\textrm{g}}$The reapplication is approved. In addition to continuing benefits from the time of adjudication, it also includes a backpay to take into account the claim history. The backpay may be accumulated with interest, the motivation being that the insurer should not have any monetary benefit from delaying the payout. The fact that payments frequently are awarded retroactively implies that the insurer might substantially under-reserve if backpay is ignored in the modeling.

${}^{\textrm{h}}$This constitutes a temporary reactivation followed by another successful reapplication resuming the benefits and likely also giving rise to another backpay related to the months for which no benefits where paid. This is once again an example of the complexity associated with adjudication of disabilities, in particular those related to mental health.

${}^{\textrm{i}}$The purpose with the vocational assessment is to identify other career choices that could allow Charlie to work more hours and hence receive less benefits. Such an assessment may be required both by the insurer and the municipality.

${}^{\textrm{j}}$This constitutes a permanent reactivation. Disability benefits typically continue until retirement, reactivation, or death, whichever occurs first.

\subsection{Interplay between agents}\label{subsec:agents}

The cases illustrate the main agents of disability insurance: the insured, the insurer, the employer, and the public system. In fact, the cases are described in a simplifying way. In the first case, about Alex's accident, an insurer was omitted from the narrative to better illustrate what would happen in the absence of insurance coverage. We may omit the insurer since the interplay between the insured and the public system is not directly affected by insurance coverage. Nonetheless, counseling and financial support can have a significant impact on the trajectory of the insured's disability. In the second case, about Charlie's claim, the role of the public system was kept exogenous. This is because this system, from an insurer's point of view, is an external environment that they and the insured are adapting to. The case is basically a description of how information about a claim arrives at the insurer, which is via direct contact with the insured and the employer rather than the public system, as well as how the insurer may respond to this information.

\begin{figure}[ht!]

    \centering

    \scalebox{1}{
    \begin{tikzpicture}[node distance=2cm and 3cm, >=stealth, thick]


    \node[draw, circle, minimum size=2.25cm] (Insured) at (1, 0) {\footnotesize{Insured}};

    \node[draw, circle, minimum size=2.25cm] (Insurer) at (7, 0) {\footnotesize{Insurer}};

    \node[draw, circle, minimum size=2.25cm] (PublicSystem) at (-2, -2) {\footnotesize{Public system}};

    \node[draw, circle, minimum size=2.25cm] (Employer) at (4, -2) {\footnotesize{Employer}};


    \draw (Insured) to (Insurer);

    \draw (Insured) to (Employer);

    \draw (Insured) to (PublicSystem);

    \draw (PublicSystem) to  (Employer);

    \draw (Employer) to (Insurer);


    \draw[gray, dashed, rounded corners] (-3.5, -3.7) rectangle (5.4, 1.4);

    \draw[gray, dotted, rounded corners] (-0.4, -3.4) rectangle (8.5, 1.7);

    \end{tikzpicture}}

    \caption{The types of agents involved in disability insurance (nodes) and their interactions (edges). Gray outlines highlight the types of agents present in Alex's case (dashed) and Charlie's case (dotted), respectively.}

    \label{fig:agents}

\end{figure}
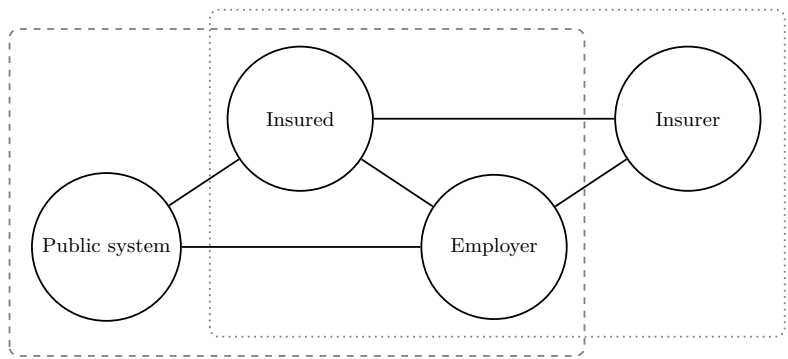

Figure~\ref{fig:agents} provides a schematic representation of the interactions between the insured, the insurer, the employer, and the public system for disability insurance, where by an interaction we refer to a direct exchange between two or more agents. One central observation is that the graph is almost fully connected, which contributes to the many-faceted complexity of the situation. The only exception is that exchange between the public system and the insurer occurs indirectly, through the insured and the insured's employer. 

It should be mentioned in this context that the insurer may interact more or less directly with the public system through, for example, political channels and in order to influence the general conditions for disability insurance, but of course this does not relate to the individual coverage and claim and is therefore not represented in the figure. The insurer may also collect additional information, for instance from public registries, which could influence the way in which they interact with the insured. In general, the insurer plays a complementary role to the public system. This characteristic also presents itself in the areas of prevention and treatment, which we discuss next.

\subsection{Prevention}\label{subsec:prevention}

Broadly speaking, insurance and prevention are two distinct and often conflicting responses to risks, and their successful integration is consequently deemed challenging~\citep{Dubois:2011}. Nonetheless, prevention has become a trendy topic in insurance, not least in regards to human health~\citep{Gauchon.etal:2020b}. This is perhaps not too surprising; the dichotomy between insurance and prevention is less pronounced for disability risk than for, say, many non-life risks. This is not least due to the fact that injury and illness, besides the economic costs that loss of earning capacity insurance might cover, has additional human and social costs -- and this reduces moral hazard; compare also with the discussion in~\citet{Botzen.etal:2019}. {In general, prevention as part of the insurance product may change the information structure of the market and, in particular, reduce the asymmetry in information between the insured and the insurer and, in turn, reduce adverse selection \citep{Peter.etal:2017}.}

In the context of disability insurance, prevention initiatives can take many different shapes and be initiated by different agents. In the following, we focus on the role of the insurer as a prevention (and not only insurance) provider. The diversity of potential disability prevention initiatives can be explored by considering various non-hierarchical taxonomies.

We can divide disability risk into two parts: frequency and severity risk. The former encompasses the probability of injury or illness causing substantial loss of earning capacity, while the latter describes the degree and duration of loss of earning capacity. There is clearly some overlap between the risks, so the division should not be considered strict. Prevention initiatives may target the frequency risk, severity risk, or both simultaneously. Sometimes, the nomenclature \textit{primary prevention} is used for an initiative that predominantly targets frequency risk, while \textit{secondary prevention} is used for an initiative that predominantly targets severity risk~\citep{Dubois:2011,Kenkel:2000}. There {are} also the concepts of \textit{self-protection} and \textit{self-insurance}, see~\citet{EhrlichBecker:1972}, which we have seen used in place of primary and secondary prevention. However, we would like to warn against such interchangeable use. Self-protection and self-insurance also include initiatives that do not directly or even indirectly reduce risk, such as risk retention. To us, prevention implies affecting or intervening in the underlying risk environment. This is fundamentally different from loss protection for the insurer due to product design or reinsurance.

The primary role of the actuary in connection with prevention initiatives lies in using mathematical and statistical methods to perform impact evaluation, that is to measure the causal effects of the initiatives. It is, after all, within the actuary's role to assess which initiatives are appropriate from a cost perspective. This necessitates, however, that the actuary is involved in identifying potential prevention areas and in the design of appropriate initiatives. Paraphrasing Ronald Fisher: To consult the actuary after the initiative is completed is merely to ask them to conduct a post mortem examination. 

As mentioned, it is possible to imagine many different prevention initiatives in connection to disability insurance. However, they all fall within one of five categories. There is the insurance coverage in itself, which provides peace of mind for the insured upon disability and consequently might have a reducing effect on the severity risk~\citep[][p.~120]{FischerKvist:2023}.  For occupational schemes, the insurer and the employer can cooperate on initiatives to improve the physical and psychological work environment and thereby reduce the frequency risk and, potentially, but likely to a lesser degree, also the severity risk. 

The remaining three categories of initiatives are of increased actuarial interest. They encompass initiatives targeting the individual insured, but at different risk stages. First, there are health promotion initiatives, such as {wearables~\citep{Spender.etal:2019,Sadowski.etal:2024}}, that aim to increase overall health and resilience. Second, prevention initiatives may be initiated just before a potential disability based on early warning indicators such as a health insurance claim or sick leave to either reduce the likelihood of disability or reduce the degree and shorten the duration. For example, the insurer may contact potential claimants and offer them additional short-term health services, such as immediate access to physical therapists and psychologists. Finally, prevention initiatives may be initialized during disability to improve the recovery rate of the insured, possibly based on pre-existing public programs. For example, the insurer might propose or even require that the insured completes a vocational assessment.

Across all categories, measuring the preventive impact of an initiative is challenging. It is necessary to separate the effect of the intervention from all other effects, including global health trends and local changes to underwriting and claims processing practices. The gold standard would be some sort of randomized trial, but this is often deemed ethically or practically undesirable. Instead, insurers might target the initiative towards high-risk individuals, adopting a cut-off selection rule. This, however, comes with additional challenges. The group subjected to the initiative would no longer be comparable to the group not subjected. Further, there is not necessarily a correlation between high-risk individuals and individuals for whom the intervention would have the largest positive effect. Therefore, causal insights are required to operationalize impact evaluations that are accurate and fit for purpose. We return to and expand on this key insight in Section~\ref{sec:prevention}.

\subsection{Systematic and unsystematic risks} \label{subsec:risks}

\begin{figure}[ht!]
    \centering
    \includegraphics[trim = 30mm 192mm 35mm 30mm, width=0.85\textwidth]{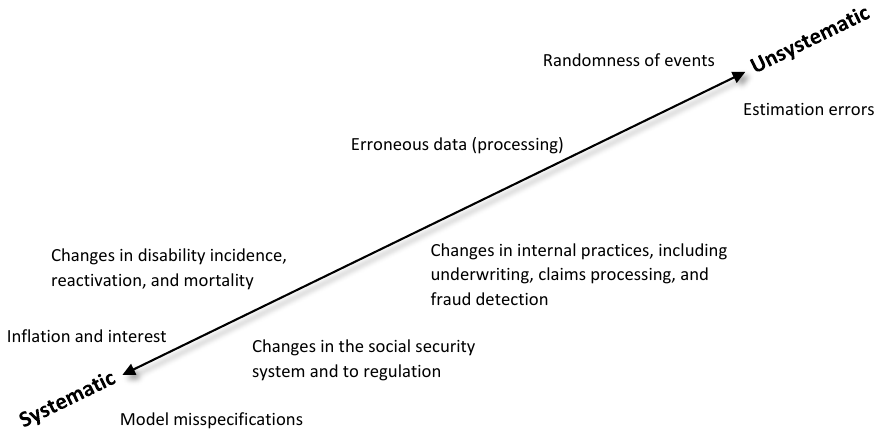}
    \caption{Primary systematic and idiosyncratic (unsystematic) risks for disability insurance.}
    \label{fig:risks}
\end{figure}

The discussion up until this point shows that disability insurance is complicated and subjects the insurer to many risks; this includes both systematic and idiosyncratic (unsystematic) types of risks. Idiosyncratic risks are specific to the insured or the insurance policy and thus manageable by diversification in accordance with the central limit theorem. Systematic risks affect the whole portfolio, or substantial segments of it, jointly and hence cannot be reduced by simply increasing the size of the portfolio. In Figure~\ref{fig:risks}, we provide an overview of the main risks associated with disability insurance.

The changes in the incidence of biometric events in particular, but also changes to internal practices and the social security system, can be abrupt, in which case one might speak of \textit{shocks}. There is, however, also significant long-term risk, so-called \textit{trend} risk. To mitigate said risk, it is essential to monitor biometric, economic, and societal trends and to adjust one's models accordingly. Reinsurance could also be an option, both in regards to biometric shocks and to counteract the interest rate risk associated with the potentially long cash flows. In general, the models should be well calibrated and supported by strong underwriting practices and fraud detection programs. Finally, prevention initiatives may reduce both unsystematic and systematic risk, confer with Subsection~\ref{subsec:prevention}.

We should like to stress that actuaries are well-positioned to contribute to the key activities involved: product design, risk modeling, and optimal prevention. In the following three sections, we address each of these areas separately. The {emphasis} will be on relevant recent developments in actuarial science and associated opportunities for actuaries of the present and the future.

\section{Product design} \label{sec:Product}

In this section, disability insurance coverages are formalized and compared with a focus on the Danish market. Our approach is initially descriptive, seeking to explore and understand the characteristics of existing products rather than discuss optimal design. Further, we are looking for common instead of distinguishing design trends, focusing on annuities. This simple-sounding task already reveals a number of challenges that hitherto have received limited to no attention in the actuarial literature.

Fundamental to the product is the specification of what constitutes lost earning capacity. The degree of lost earning capacity is usually taken to be the proportion of hours the insured is able to work compared to a standard number of working hours. The proportion may be based on the hours it is possible to work in the insured's own original occupation, a similar occupation, or any occupation. These specifications play an important role not least because they may influence the incidence rate, reactivation rate, and payment sizes considerably, which also has implications for subsequent actuarial modeling. Due to the rich interplay with the public system, it is often natural for the definition stipulated in the insurance contract to be at least partly aligned with the one used in the public system, which may result in the insurance coverages being subjected to the whims of politics. In Denmark, three-fifths of the insured receiving disability benefits in 2020 were awarded benefits based on eligibility for early retirement pension, while the remaining two-fifths were awarded based on the insurer's internal health assessment of the insured~\citep[][Boks~6.4]{Ekspertgruppe:2022}.

Having specified what constitutes loss of earning capacity, the product further depends on the specification of the size of the compensation in case of a disability. In Subsection~\ref{subsec:ContractualPay}, realistic contractual payments are formalized. Contractual payments are those that are stipulated in the insurance conditions. Further complexity arises in ensuring that the insured receives the benefits they were eligible for due to reporting delays and adjudications; this is the topic of Subsection~\ref{subsec:ClaimsettlePay}.

\subsection{Contractual payments} \label{subsec:ContractualPay}

We here give a continuous time description of the contractual payments. This is a natural approach since the underlying biometric events occur in continuous time. In reality, payment streams feature lump-sum payments at discrete points in time, but payment rates are mathematically convenient and offer a useful approximation when the payment frequency is sufficiently high; this is the case for disability annuities with monthly payments. 

\subsubsection{Simple disability annuity} \label{subsubsec:simpleDisab}

\begin{figure}[ht!]
    \centering
    \scalebox{0.85}{
    \begin{tikzpicture}[node distance=8em, auto]
	\node[punkt] (i1) {Active};
        \node[anchor=north east, at=(i1.north east)]{$1$};
        \node[punkt, right=3cm of i1] (i2) {Disabled};
        \node[anchor=north east, at=(i2.north east)]{$2$};
        \node[, right=1.5cm of i1] (a) {};
        \node[punkt, below=1.5cm of a] (a2) {Dead};
        \node[anchor=north east, at=(a2.north east)]{$3$};
        \path (i1) edge [pil, bend right=15] node [below=0.15cm]  {} (i2)
	;
         \path (i2) edge [pil, bend right=15] node [above=0.1cm]  {} (i1)
	;
         \path (i1) edge [pil] node [left=0.15cm]  {} (a2)
	;
         \path (i2) edge [pil] node [right=0.15cm]  {} (a2)
	;
    \end{tikzpicture}}
    \caption{State space $\mathcal{J}$ for a classic disability model. The arrows represent the possible transitions.}
    \label{fig:ClassicDisabMultistate}
\end{figure}
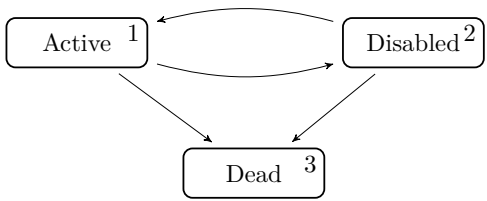

Multistate models provide a parsimonious way to formalize the contractual payments. Let $Y=\{Y(t)\}_{t \geq 0}$ be a stochastic process that is piecewise constant and which takes values in the state space $\mathcal{J}$ depicted in Figure~\ref{fig:ClassicDisabMultistate}. The definition of \textit{Disabled} in the model is to be understood as being eligible for disability benefits according to the stipulations in the insurance conditions, and the contractual payments $B=\{B(t)\}_{t \geq 0}$ of a simple disability annuity may then be cast as
\begin{align*}
    \diff B(t) = 1\{Y(t)=2\} b \diff t,  \hspace{0.75cm}  B(0)=\pi,
\end{align*}
where $\pi < 0$ is the initial premium and $b > 0$ is the disability benefit rate. This definition and corresponding mathematical expression hide the stipulations for eligibility. The latter usually require the disability to have occurred within the coverage period, which commonly spans one to three years, the proportion of lost earning capacity to be sufficiently high, for example 50 \%, and a deferred (waiting) period of, say, three months to have passed.  

Defining the disabled state as being eligible for disability benefits has the effect of placing the difficulty in modeling onto the probabilities of transitioning to and from the disabled state, rather than placing it in the payment rates. This implies, among other things, that the time of disablement and the transition probabilities become product dependent. However, using a product-independent definition of the time of disablement such as the first day of sickness leads to other complications in formalizing the contractual payments. Taking the deferred period as an example, one could add an indicator to the benefits that the duration in the disabled state has exceeded the deferred period, but this might not correctly describe what happens in practice since the deferred period is often annulled if the insured relapses after a short reactivation. 

\begin{figure}[H] 
	\centering
	\scalebox{0.85}{
	\begin{tikzpicture}[node distance=2em and 0em]
		\node[punktl] (2) {Reactivated};
		\node[anchor=north east, at=(2.north east)]{$3$};
		\node[punkt, left = 20mm of 2] (1) {Disabled\,};
		\node[anchor=north east, at=(1.north east)]{$2$};
		\node[draw = none, fill = none, left = 18 mm of 2] (test) {};
		\node[punkt, left = 20mm of 1] (3) {Active};
		\node[anchor=north east, at=(3.north east)]{$1$};
		\node[punkt, below = 15mm of 1] (4) {Dead};
		\node[anchor=north east, at=(4.north east)]{$4$};
	\path
		(1)	edge [pil]				node [above]			{}		(2)
		(3)	edge [pil]				node [above]			{}		(1)
	   (1)	edge [pil]				node [above]			{}		(4)
	   (2)	edge [pil]				node [above]			{}		(4)
       (3)	edge [pil]				node [above]			{}		(4)
	;
	\end{tikzpicture}}
	\caption{State space for an alternative disability model with separate reactivated state. The arrows represent the possible transitions.}
	\label{fig:AltDisab}
\end{figure}
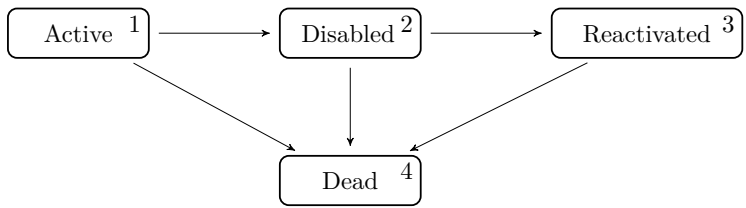

Thus, even in this simple setting where benefits are constant and claim settlement processes are not involved, the non-hierarchical nature of disability trajectories leads to disability insurance having a high level of complexity compared to other types of coverages. These complexities seem to be ignored in the classic literature on actuarial multistate modeling. In the formulations of Example~3.2 in~\citet{Christiansen:2012} and Section~3.2.2 in~\citet{Haberman:Pitacco:1998}, it is for instance not possible to receive payments when entering the disabled state after the end of the coverage period and the deferred period resets after each entry to the disabled state. This means that temporary reactivations for the same underlying disability event are not accounted for. However, if temporary reactivations or multiple distinct disabilities are disregarded, the added complexities disappear, and one may thus give an approximate description of the contractual payments based on the classic literature by ignoring multiple disabilities and using the alternative state space depicted in Figure~\ref{fig:AltDisab}.

When disability benefits are constant, underwriters should manually take people's financial situation into account and anticipate what benefits they will receive from other sources such as the public system. If the insured's total income would increase substantially in the event of a disability, as they may receive income from several sources, there is the potential for moral hazard as noted in Chapter 3 of~\citet{Haberman:Pitacco:1998}, but this effect is likely limited due to the reasons outlined in Subsection~\ref{subsec:prevention}. Another downside, however, is that the insured has potentially over-insured themselves and therefore pays a premium that is too high in relation to the utility value of the coverage. On the other hand, if the insured's income substantially decreases in the event of a disability, then the insurance is not achieving its purpose of ensuring that the individual can maintain a similar or only slightly lower level of consumption. Accommodating these complications requires tailored insurance coverages since the public benefits change as the disability progresses. Such complicated coverages designed around the social security system, which are by now rather common, are the focal point of the next subsection. 

\subsubsection{Disability annuity} \label{subsubsec:disab}

Using the state space from Figure~\ref{fig:ClassicDisabMultistate} with the same definition of the \textit{Disabled} state, a simple disability insurance product designed around the social security system has contractual payments on the form
\begin{align*}
\diff B(t) = 1\{Y(t)=2\} \min\{  \max \{ s-e_t-c_t , 0 \} , d \times s  \} \diff t, \hspace{0.75cm}  B(0)=\pi,
\end{align*}
where $d$ is the percentage of salary covered, for example two-thirds, $s$ is the salary just prior to the disability, $e_t$ is the earning capacity at time $t$, and $c_t$ is the compensation from social security and other insurance coverages at time $t$. 

The Danish market also features more complicated coverages. Inflation regulation is, for instance, a common feature; here $s$ is scaled with an appropriate inflation index as to maintain the purchasing power. Additionally, the deduction of benefits from other insurance coverages may first take place when the total income exceeds some higher proportion of the previous salary than $d$, say 80 \%. The benefit rate might also be reduced whenever the insured is enrolled in public benefit programs that deduct compensation from insurance coverages; an example hereof is the resource clarification program. Furthermore, there is the entanglement with tax rules, pension contributions, and labor market contributions. The intention behind such modifications to the simple product is to stabilize the insured's total income level, while keeping the insurer's expenses -- and therefore the premium rate -- as low as possible. {In addition to the dependence on public benefit programs, there might also be differing levels of benefits according to the type or severity of disability. In principle, all these features can be handled by expanding the disability state, confer with~\citet{Sandqvist:2023}, but with a substantial increase in the dimensionality of the problem as a consequence.}

These types of products are not well-studied in the actuarial literature. However, they can be studied using well-known methods from event history analysis and the mathematics of life insurance. It is worthwhile to note that it is not necessary to explicitly forecast $e_t$ and $c_t${, for example by expanding the disability state,} to calculate reserves and premiums; it suffices to model the average payments at future times (by the law of iterated expectations). This way of thinking might also be applied to incorporate some of the complexities described in Subsection~\ref{subsubsec:simpleDisab}, rather than just placing it on the transition probabilities. The actuarial community would benefit from more work in these areas.

\subsection{Claim settlement payments} \label{subsec:ClaimsettlePay}

While the insurance contract determines the payments that the insured is eligible to, as described in Subsection~\ref{subsec:ContractualPay}, the actual process of awarding the insured these benefits is complicated by the claim application process. 

There is both a reporting delay incurred by the insured (from the occurrence of the event until the insured notifies the insure{r}) and a further delay incurred by the insurer due to its adjudication process (from when the insurer is notified until eligibility is determined). If the insurer determines that the insured is eligible for benefits, the insurer has to ensure that the insured is compensated according to the stipulations in the insurance contract. This is done by awarding backpay from the end of the deferred period to the time where the decision to award benefits is made -- in addition to the annuity payments from that time until the insured reactivates, dies, or reaches their retirement age. In determining the size of backpay, the time value of money has to be taken into account. Therefore, backpay is usually accumulated with interest matching the periods that the benefits pertain to. A similar phenomenon occurs if benefits are stopped by the insurer, but the insured reapplies and successfully demonstrates that the payments were wrongfully stopped. The realized cash flow $\mathcal{B}=\{\mathcal{B}(t)\}_{t\geq0}$ for the simple disability annuity of Subsection~\ref{subsubsec:simpleDisab} consequently takes the form
\begin{align*}
\diff \mathcal{B}(t) = 1\{Z(t)=2\} b \diff t + \textnormal{backpay}(t) \diff N(t)
\end{align*}
where $Z$ is a stochastic process taking values in the same space as $Y$, but only visiting the state \textit{Disabled} when the insured is actually receiving annuity benefits, and where $N(t)$ counts the number of times backpay has been awarded before time $t$. The backpay takes the form 
\begin{align*}
\textnormal{backpay}(t) = \int_{\alpha(t)}^{\beta(t)} \exp\left(\int_s^t r(v) \diff v\right) b \diff s
\end{align*}
with $\alpha(t)$ and $\beta(t)$ delimiting the relevant period and $r$ denoting the interest rate.

These complications have only recently been addressed in the multistate modeling literature on disability insurance, namely in~\citet{Buchardt.etal:2023a, Buchardt.etal:2023b} and~\citet{Sandqvist:2023}. One may therefore reasonably ask whether they can be ignored. For pricing or reserving at the inception of the contract, the answer is to a large degree `yes', but for risk management and reserving during and after the coverage period, the answer turns out to be a resounding `no' unless the reporting and adjudication delays are very short, see Subsection~\ref{subsec:multistateModels} for detailed explanations as to why. 

In the context of estimation, the presence of reporting delays as well as incomplete adjudications biases the sample. Fitting a model to have good predictive performance on the observed data, for instance using out-of-sample-based methods, leads to a biased model -- because the sample is biased. Reporting delays result in the sample containing too few observed disabilities, while later rejections upon adjudication result in the sample containing too many observed disabilities. It is therefore not even clear whether one over- or underestimates the disability incidence rate. The actuary plays an important role in recognizing and taking these effects into account when performing reserving and impact evaluations for disability insurance. In fact, actuarial reserving as a field has in many ways evolved around accommodating such sampling effects, which in addition to reporting delays and incomplete adjudication encompasses left-truncation and right-censoring stemming from the finite observation window and from insured entering and leaving the portfolio.  

\section{Actuarial modeling} \label{sec:Modeling}

As seen in Section~\ref{sec:Product}, formalizing the contractual payments for a single insured is inherently difficult, and this also complicates the risk modeling. In this section, we discuss and compare relevant actuarial models, providing new insights into the adequacy of current stochastic modeling approaches when applied to disability insurance coverages. The overall aim is to obtain sound actuarial models for risk management, while maintaining as much simplicity as possible; they should have predictive prowess, and it should be possible to validate them in- and out-of-sample. This aim epitomizes the identity and role of the actuary since achieving it requires a synthesis of insight about the design of the products, the associated biometric and financial risks, as well as the available data and its consequences for the probabilistic and statistical aspects of the model.   

\subsection{Reserving and pricing}

Premiums are said to be actuarially fair if they equal the expected expenses. From a narrow mathematical standpoint, obtaining actuarially fair premiums is therefore a special case of reserving. Reserves represent the value of the future cash flow, with market-consistent valuation leading to reserves $V$ on the form
\begin{align} \label{eq:reserve}
    V(t) &= \mathbb{E}\left[ P(t) \mid \mathcal{F}_t \right], \\
    \label{eq:presentValue}
    P(t) &= \int_t^\infty \exp\left(-\int_t^s r(v) \diff v\right) \diff B(s),
\end{align}
where $r$ is the interest rate, $\diff B$ is the cash flow, $\mathcal{F}_t$ is the information that is known at time $t$, and time is measured as the duration since the inception of the contract. This is the (conditional) expectation of the present value of future benefits with respect to some market-consistent (equivalent) probability measure. Actuarially fair premiums are those for which $V(0) + B(0) = 0$, which in our setting implies an initial premium $\pi = -V(0)$. 

{ There are well-established methods available for computing reserves when the following three conditions are met: policies are independent, the cash flow can be written in a (semi-)Markov form, and the cash flow is independent of the interest rate; see for example~\citet{Norberg:1991},~\citet{Milbrodt:Stracke:1997},~\citet{Buchardt.etal:2015},~\citet{Bladt.etal:2020},~\citet{Christiansen:2021}, and~\citet{Ahmad.etal:2023}. As discussed in Subsections~\ref{subsec:ContractualPay} and ~\ref{subsec:ClaimsettlePay}, it is not always straightforward or even possible to write the cash flow of a disability insurance product in a (semi-)Markov form, and the realized cash flow is typically even more complicated. Furthermore, the independence assumption may be violated if the frequency of events is affected by economic conditions, including individuals being less likely to reactivate during economics downturn due to increased unemployment, or if the payments themselves depend on economic conditions, for example if they are linked to salary or inflation. Literature relevant to these challenges is discussed in Subsection~\ref{subsec:multistateModels}.}   

{ Even though pricing may be a special case of reserving, in practice they usually have different goals.} For pricing, it is important to have precise premium sizes on an individual level to avoid adverse selection. For reserving, both accurate \textit{size} and \textit{timing} are important. The insurer has to charge enough money in the beginning to pay the later claims, which favors prices that are accurate on a portfolio level. Regulation compels the insurer to set aside additional capital to maintain solvency even in cases that are considerably worse than current best estimates, which may be thought of as accounting for the market price of risk. These are aspects related to the size of the reserve. To have good timing, the reserve has to be close to the present value of future benefits at each point over the duration of the contract period, not just at the beginning. Otherwise, there will be a mismatch between the release of reserves and incoming losses. This again leads to swings in the insurer's financial results and uncertainty about the insurer's financial situation that makes it difficult to run an efficient operation. Furthermore, the timing of the reserve is crucial for hedging of the financial risks associated with the future cash flow. If possible, the insurer would also want good timing on a more granular level to be able to monitor the business on a sub-portfolio level. 

In addition to size and timing, actuaries also care about \textit{model complexity} and \textit{statistical complexity}. Model complexity comes in many forms. We here restrict our attention to the complexity of the mathematical concepts involved in specifying the model, and not for example the explainability of the model and its output. This notion of model complexity is relevant for the time and skill needed to understand the model, but also influences the risk of implementation errors and the ease with which the model may be communicated to non-experts. Statistical complexity is taken to include the hardness of both the statistical learning problem and the practical issues of modeling choices and implementation. In the following three subsections, different model paradigms are discussed with these considerations in mind; a comparison may be found in Subsection~\ref{subsec:recap}.

\subsection{Non-life individual reserving models} \label{subsec:nonlife}

Since formalizing the contractual payments is difficult, one may ask whether it is more favorable to disregard the known structure of the cash flow $\diff B$ and also learn this from the data. This is similar to the individual reserving approach from non-life insurance, and one could hence use recent methods from this area such as~\citet{Crevecoeur.etal:2022b} or~\citet{Yang.etal:2024}. Their approach is to model the full real-time development of the claim and all time-varying covariates; note, however, that~\citet{Yang.etal:2024} assume the entire path of the time-varying covariates to be known. A brief summary goes as follows. A high-dimensional stochastic process $X=\{X(t)\}_{t\geq0}$, which represents all available policy and claim level information, is introduced, and the available information is specified as $\mathcal{F}_t = \sigma\{X(s) : s \leq t \}$. The distinction between contractual and realized payments is abandoned, meaning $\diff B$ is taken to just be $\diff \mathcal{B}$; compare to Subsections~\ref{subsec:ContractualPay} and~\ref{subsec:ClaimsettlePay}. This choice implies that $B(t)$ is recoverable from the information $\mathcal{F}_t$, so the present value in~\eqref{eq:presentValue} is indeed the relevant present value, and the reserve of~\eqref{eq:reserve} is operational. However, the reserve does not admit a closed-form expression, and since calculating them numerically via differential equations is undesirable due to the high dimensionality of the model, Monte Carlo methods are used.

This approach makes the probabilistic description of the model relatively simple, but results in a learning problem with high statistical and computational complexity. In particular, the curse of dimensionality implies that the appropriate amount of data needed to estimate the model grows quickly with the dimensionality of the model, not least since the signal-to-noise ratio is low in insurance applications. { Machine learning models have recently become popular for individual reserving problems as they handle  high-dimensional and non-linear data more effectively, see for example~\citet{Wuthrich:2018a},~\citet{Delong.etal:2021},~\citet{Bucher:Rosenstock:2024},~\citet{Holvoet.etal:2025}, and~\citet{Avanzi.etal:2026}. While often preferable to restrictive parametric models in such a setting, they are still susceptible to the curse of dimensionality and other fundamental limitations of statistical inference~\citep{Stone:1980,Stone:1982,Vapnik:1998}. Furthermore, they are susceptible to whatever structural assumptions are imposed on the model such as the influence of reporting delays and the independence between individuals, claims sizes, and claims frequency. Perhaps it is for this reason that machine learning has yet had limited success for disability insurance -- due to structural and sampling issues. For concrete examples of such structural assumptions, note that the Poisson assumption of~\citet{Crevecoeur.etal:2022b} would be unpleasant for disability insurance applications since one can in reality only be compensated for at most one disability at any time. Similarly, it is in~\citet{Yang.etal:2024} assumed that} the different payment sizes and waiting times are independent given the settlement time, which does not describe disability annuities well since payments occur monthly after the first payment and the payment sizes are similar for payments that are temporally close. 

Another challenge with the non-life approach is that the individual model components are tuned to be accurate for their separate regression problems, and this may lead to substantial bias for the aggregate reserves due to plug-in bias. In~\citet{Crevecoeur.etal:2022a}, it is suggested to remove this aggregate bias by rescaling the predictions such that the aggregate predictions historically agree with the aggregate observations. An alternative that does not seem to have been explored in an actuarial context, but which might be useful, is to remove the plug-in bias for the portfolio reserve using a so-called one-step estimator, confer with Section~4 of~\citet{Kennedy:2022}.   

\subsection{Aggregate reserving models}

\subsubsection{General considerations}

A different approach, which also does not utilize the known structure of the contractual payments, is to use aggregate models for which $\diff B$ represents the realized cash flow of the portfolio instead of a single individual, and for which the information used for reserving, written $\mathcal{F}$, is kept at portfolio level (even if more granular information is available). Aggregate models have the advantage that they target the portfolio level reserve and that they are comparatively simple to implement. 

However, aggregate models have four general disadvantages. First, they cannot use granular data, which may impede their predictive performance. Second, they are slow to capture shifts in the covariate composition of the portfolio or new trends such as an increase in mental health-related claims. Third, they lack explainability in the sense that the models offer little assistance in pinpointing underlying reasons for deviations between observations and predictions. Together, these three points make it difficult to monitor the business on a more granular level and to make ad-hoc adjustments based on additional external expert information. Fourth, they do not provide consistent estimates when used to impute right-censored claims in order to obtain a regression sample for fitting pricing models as is often done in non-life insurance; see the discussion in~\citet{Crevecoeur.etal:2022b}, especially Section~3.1. 

\subsubsection{Application to disability insurance}

There are also some additional challenges associated with using aggregate models for disability insurance reserving. The aggregate reserving literature has generally evolved around the assumption that the available data consists of payments occurring in a run-off triangle consisting of accident years (rows) and development years (columns). The run-off triangle is commonly, but not always, assumed to be aggregated on some time grid, say monthly or yearly. Since disabilities frequently lead to several decades of benefits, while the eligibility criteria change rather frequently, one has in most situations too few years of data to be able to estimate the outstanding liabilities using chain-ladder, or variants thereof, since no complete run-off is observed. It is hence almost always necessary to impose some structural assumptions that allows one to extrapolate the future expenses for late development years. 

Individual-level data shows that insured who have received benefits for a couple of years have very low reactivation and mortality rates, so claims that are open after a couple of years generally run until the retirement age. Furthermore, in Denmark, the social security benefits also tend to stabilize after some years since insured who remain disabled find a suitable flex job or retire early. Therefore, a pragmatic approach could be to first model the average age of insured who have open claims after some fixed number of years, and then in a second step to cast subsequent payments as constant until the retirement age subtracted the aforementioned average age. This would likely lead to a reasonable size of the reserves, but poor timing.

If data on individual ages is available, a perhaps preferable alternative is to include age as a covariate in the aggregate model. Some approaches for incorporating such individual information in chain-ladder models are given in~\citet{Wuthrich:2018a} and~\citet{Delong.etal:2022}. It also seems like it would be possible to extend the approach of~\citet{Bischofberger.etal:2020} to include covariate information such as age, but the setup comes with other strong assumptions that are undesirable for modeling disability insurance, namely that payments are independent and all claims consist of single payments. If portfolio reserving is the only objective, we expect that the best size and timing can be obtained by leveraging granular data, but using a model that targets the portfolio reserve as discussed here and at the end of Subsection~\ref{subsec:nonlife}. This, however, comes at the cost of a significant increase in complexity compared to traditional aggregate models. 

\subsection{Multistate individual reserving models} \label{subsec:multistateModels}

\subsubsection{Literature and challenges} \label{subsubsec:literature}

The actuarial literature on individual disability modeling using multistate methodology is substantial, confer with~\citet{Taylor:1971},~\citet{Segerer:1993},~\citet{Haberman:Pitacco:1998},~\citet{Christiansen:2012},~\citet{Sandqvist:2023}, and the references therein. {The introduction of an economic-demographic environment may furthermore lead to tractable dependence between individuals, confer with~\citet{Djehiche:Lofdahl:2014,Djehiche:Lofdahl:2018} and~\citet{Aro.etal:2015}; in~\citet{Djehiche:Lofdahl:2014}, the underlying risk factors are assumed to observable, while they are unobserved (latent) in the subsequent works.} Nonetheless, the increasing complexity outlined in Subsection~\ref{subsubsec:simpleDisab} arising from temporary reactivations and several distinct disabilities has only just received some awareness, see~\citet{Sandqvist:2023}, and the situation where payments are on the form described in Subsection~\ref{subsubsec:disab} also warrants further exploration. The latter situation can to some extent be compared to the situation where the payment rate depends on the degree of disability as in Subsection~4.2 of~\citet{Segerer:1993}, so that the annuity formula is scaled with the mean degree of disability. Further, this approach can be extended to model mean benefits in the face of deduction from social security benefits, confer with Appendix~A.1 and~A.2 of~\citet{Sandqvist:2023}. We believe that this is an important avenue for actuaries and actuarial scientists to explore.

There is also the effect of the claim settlement process to take into account, confer with Subsection~\ref{subsec:ClaimsettlePay}. The implications of reporting delays on the estimation of disability incidents has been studied in~\citet{Kaminsky:1987},~\citet{Konig.etal:2011}, and~\citet{Aro.etal:2015}. In the context of reserving, it is noted in~\citet{Segerer:1993} that insurers can set up IBNR reserves if they have the data to support them, but additional details are not provided. It is only recently in~\citet{Buchardt.etal:2023a} and~\citet{Sandqvist:2023} that the problem has been attempted to be tackled in a somewhat general manner. The topic is, however, frequently explored in non-life insurance contexts with approaches similar to those described in Subsection~\ref{subsec:nonlife}, see for example~\citet{Norberg:1993,Norberg:1999},~\citet{Haastrup:Arjas:1996},~\citet{Antonio:Plat:2014},~\citet{Lopez.etal:2019},~\citet{Okine.etal:2022},~\citet{Crevecoeur.etal:2022a,Crevecoeur.etal:2022b}, and~\citet{Bucher:Rosenstock:2024}.

The classic multistate approaches to disability insurance modeling have focused on reserves on the form~\eqref{eq:reserve} with the specifications that $\diff B$ are the contractual payments and the information $\mathcal{F}_t$ equals $\sigma\{ Y(s) : s \leq t\}$ for the biometric state process $Y$ generating the contractual payments. This leads to two fundamental issues for reserving whenever there are non-negligible reporting delays and adjudication processes, and these challenges do not seem to have been discussed prior to~\citet{Buchardt.etal:2023a} and~\citet{Sandqvist:2023}. The issues are as follows. First, if there are reporting delays and adjudication processes then the choice $\mathcal{F}_t=\sigma\{ Y(s) : s \leq t\}$ is not operational for reserving since this information might not be available at time $t$. For example, the insured can become disabled before time $t$, but report this after time $t$. Similarly, since $\{Y(s) : s \leq t\}$ might not have been observed at time $t$, the insurer might not have paid $B(t)$ at time $t$, so the present value in~\eqref{eq:presentValue} is not the relevant present value. For example, the insurer might not have paid any disability benefits by time $t$, while the insured is disabled and eligible for benefits before time $t$, which would then lead to backpay after time $t$. 

\subsubsection{Recent contributions}

Recently,~\citet{Buchardt.etal:2023a} and~\citet{Sandqvist:2023} introduced a multistate framework that seeks to accommodate the challenges outlined in Subsection~\ref{subsubsec:literature}. The idea is to keep $V$ defined as in the classic multistate modeling framework, but introduce another reserve $\mathcal{V}$ according to
\begin{align*}
\mathcal{V}(t) &= \mathbb{E}\left[ \mathcal{P}(t) \mid \mathcal{G}_t \right], \\
\mathcal{P}(t) &= \int_t^\infty \exp\left(-\int_t^s r(v) \diff v\right) \diff \mathcal{B}(s),
\end{align*}
where $\diff\mathcal{B}$ is the realized cash flow and $\mathcal{G}$ is the available policy- and claim-level information. In other words, this is essentially the same reserve as used in non-life insurance contexts, confer with Subsection~\ref{subsec:nonlife}, and hence uses the correct present value with an operational information. The modeling paradigm, however, deviates from the non-life approach by retaining the a priori known structure of $\diff B$ and introducing relevant structural assumptions regarding the relation between $\diff B$ and $\diff\mathcal{B}$ as well as between $\mathcal{F}$ and $\mathcal{G}$. These assumptions allow one to express $\mathcal{V}$ in terms of the classic reserve $V$ with certain natural modifications.

In particular, as briefly mentioned at the end of Section~\ref{sec:Product}, one has $\mathcal{V}(0)=V(0)$ since $\mathcal{G}_0=\mathcal{F}_0$ and $\mathcal{P}(0) = P(0)$. The former is because at policy initiation the state of the insured is known and no additional claim settlement information is available, while the latter follows from the fact that no backpay may relate to events before policy initiation. Therefore, the effect of the claim settlement process may be ignored at policy initiation. However, if there are significant reporting and adjudication delays, then there may be substantial discrepancies between the two filtrations and the two present values after time $t$. This cannot be ignored if the reserves are to have good timing.

A disadvantage of this modeling paradigm is that it usually necessitates rather strong independence assumptions which if violated may impair both the size and timing of the reserves. It is possible to accommodate certain violations of the assumptions, confer with Remark~3.9 and~3.10 of~\citet{Sandqvist:2023}, but only at the cost of additional model complexity. Another disadvantage is the need for quite granular data to be available to the actuary.

The statistical and computational complexity is comparable to current state-of-the-art models in the multistate life insurance literature and therefore moderate. Furthermore, the probabilistic description of the model is rather complicated, but not substantially more demanding than for regular multistate models. This means that if one is already using multistate models for reserving, it can be considerably easier to adopt the methodology of~\citet{Buchardt.etal:2023a} and~\citet{Sandqvist:2023} compared to switching to a model inspired by non-life insurance methods. This may be of particular interest to insurers based in Denmark, not least due to the rich tradition for multistate modeling as reflected in the literature emanating from Copenhagen.

In comparing the multistate paradigm to the non-life individual paradigm described in Subsection~\ref{subsec:nonlife}, advantages are found to be that the structure of the contractual payments is exploited, the dimension of the model is likely considerably lower, and that the reserve may be calculated efficiently via differential equations rather than Monte Carlo methods. Furthermore, the resulting reserve is composed of modifications of classic reserves and hence easier to interpret and adjust on an ad-hoc basis. Based on Theorem~3.4 in~\citet{Sandqvist:2023}, the reserve for an insured with no reported disability claim is approximately
\begin{align*}
\mathcal{V}(t) = V_a(t) + \int_0^t V_i(s,0) \mathbb{P}(\textnormal{Reporting delay} > t-s) \mu_{ai}(s) \diff s,
\end{align*}
where $V_a(t)$ is the classic reserve for an insured that is active at time $t$, $V_i(s,u)$ is the classic reserve for an insured that became disabled at time $s-u$, and $\mu_{ai}$ is the disability incidence rate. The first term corresponds to disabilities that are covered but not incurred, while the second term corresponds to disabilities that have occurred but have yet to be reported.

The situation of a reported disability claim is more involved, since one m{u}st distinguish between three cases: reported-but-not-paid, currently eligible for disability benefits, and previously eligible for disability benefits. Given the information available at time $t$, let $G(t)$ denote the time point from which the insured is eligible for disability benefits if the claim is (re)awarded and let $W(t)$ be the corresponding disability duration at $G(t)$. The reported-but-not-paid reserve then approximately reads
\begin{align*}
\mathcal{V}(t) = \mathbb{P}(\textnormal{Claim is awarded} \mid \mathcal{G}_t ) V_i(G(t),0).
\end{align*}
This follows from Theorem~3.7 in~\citet{Sandqvist:2023}. Furthermore, the reserve for an insured currently eligible for disability benefits is
\begin{align*}
\mathcal{V}(t) = V_i(t,W(t)),
\end{align*}
confer with Theorem~3.8 in~\citet{Sandqvist:2023}. Finally, the reserve for an insured previously eligible for disability benefits is approximately
\begin{align*}
\mathcal{V}(t) = \mathbb{P}(\textnormal{Claim is reawarded} \mid \mathcal{G}_t ) V_i(G(t),W(t)).
\end{align*}
For estimation of the parameters required to calculate $\mathcal{V}$, in~\citet{Sandqvist:2023} it is established that this is a special case of the statistical problem studied in~\citet{Buchardt.etal:2023b}, so that their methods and results may be directly applied. Subject to certain simplifications, this leads to the following solution for the estimation of the disability and reactivation hazards. The exposures $E(t_i)=\int_{t_i}^{t_{i+1}} 1\{Y(s)=1\} \diff s$ for the disability hazard have to be multiplied by $\mathbb{P}(\textnormal{Reporting delay} \leq  t-t_i)$, where $t$ is now the time of statistical analysis. Furthermore, reported-but-not-paid occurrences qualify as disability occurrences, but only after scaling with $\mathbb{P}(\textnormal{Claim is awarded} \mid \mathcal{G}_t )$, while (temporary) terminations of disability benefits qualify as (permanent) reactivations, but only after scaling with $1- \mathbb{P}(\textnormal{Claim is reawarded} \mid \mathcal{G}_t )$. Finally, the scaling factors { $\mathbb{P}(\textnormal{Reporting delay} > t-s)$, $\mathbb{P}(\textnormal{Claim is awarded} \mid \mathcal{G}_t )$, and $\mathbb{P}(\textnormal{Claim is reawarded} \mid \mathcal{G}_t )$} can be estimated by somewhat standard event history analysis methods. Altogether, this gives rise to a two-step estimation procedure in which estimation of the scaling factors precedes estimation of the hazards.
\subsection{Comparison} \label{subsec:recap}

Table~\ref{tab:recap} offers a summary of the points discussed in Subsections~\ref{subsec:nonlife}--\ref{subsec:multistateModels}. It is not intended to capture all nuances, but rather to indicate the expected relative potential of the different modeling paradigms. For the prediction column, the entries represent the performance on a portfolio and individual level, respectively. Other aspects discussed but not represented in the table are explainability and computational complexity. It is up to the actuary to determine which modeling paradigms is best suited for risk assessment and management, including pricing and reserving, taking into account the resources, challenges, and strategies that are pertinent to their situation.

\begin{table}[ht!]
    \centering
    \scalebox{0.8}{
    \renewcommand{\arraystretch}{1.3} 
    \begin{tabular}{ p{2cm} |  >{\centering\arraybackslash}p{3cm} >{\centering\arraybackslash}p{3cm} | >{\centering\arraybackslash}p{2.5cm} >{\centering\arraybackslash}p{2.5cm} }
        
         & \multicolumn{2}{c|}{\textbf{Prediction}} & \multicolumn{2}{c}{\textbf{Complexity}} \\
          
         & \textbf{Size} & \textbf{Timing} & \textbf{Model} & \textbf{Statistical} \\
        \hline
        \textbf{Non-life} & poor/good & poor/good & moderate & high \\
        
        \textbf{Aggregate} & decent/poor & decent/poor & low & low \\
        
        \textbf{Multistate} & decent/decent & decent/good & high & moderate \\
        
    \end{tabular}
    }
    \caption{Characteristics of the different actuarial modeling paradigms: non-life individual reserving models, aggregate reserving models, and multistate individual reserving models. The entries for size and timing are on the form portfolio/individual performance. For example, aggregate models are expected to yield reserves with decent size on a portfolio level and poor size on an individual level.}
    \label{tab:recap}
\end{table}

\section{Impact evaluation for prevention} \label{sec:prevention}

While prevention is a topical subject in insurance, there does not exist much actuarial literature dedicated to it. Perhaps this can be explained by the fact that actuaries need not play a prominent role whenever a prevention initiative merely constitutes an implementation of existing principles based on empirical evidence from, for instance, the health sciences. However, the situation of insurers' often differs from those in the health sciences literature: They are a different kind of health partner than, for example, an employer or care provider; their target demographic may differ from, for example, the wider population; and their basis and options for action may look different. It is therefore important to be aware that empirical insights from the literature cannot necessarily be applied one-to-one to the insurer's situation. Consequently, insurers would benefit greatly from being able to quantify the impact of prevention and treatment initiatives using their own data -- in combination with theoretical as well as empirical insights from the broader scientific literature.

Both portfolio-level and more granular impact evaluations are valuable. A portfolio-level impact evaluation can reveal whether an initiative in force is cost effective, in the sense that the health promotion (measured as a reduction in disability benefit payout) exceeds the costs of administering the initiative. A more granular impact evaluation, on the other hand, could provide valuable feedback regarding the optimal design of the initiative and its primary target group.

Among the few actuarial papers dedicated to prevention, we highlight~\citet{Gauchon.etal:2020a,Gauchon.etal:2020b,Gauchon.etal:2021}. In~\cite{Gauchon.etal:2020b,Gauchon.etal:2021}, a ruin theoretic setup is considered and the optimal amount to invest into prevention is studied with prescience about how changes in investment affect the rate of claim arrivals. To operationalize this theoretic contribution, a natural question is: \textit{How can one infer the effect of investments into prevention on the claim arrivals and the claims sizes?} Furthermore, since the same monetary amount could be invested in many different ways, further relevant question are: \textit{Which initiatives are most effective in reducing the risks? Are these effects heterogeneous across different groups of insured -- and in which sense?} 

In~\citet{Gauchon.etal:2020a} and to address the second and third question, it is proposed to cluster insured in terms of their consumption of different health services, for example psychiatry and radiology consultation, and then target prevention programs  to high-risk individuals. This is a good starting point: if the risk is high, it may be easier to achieve a substantial risk reduction. However, this might not be a cost-effective solution. For example, insured at moderate risk could perhaps be relatively more receptive to the prevention initiative. There are clear parallels to the following classic case from sales:
\begin{center}
\begin{minipage}{.8\textwidth}
\textit{Consider a department of telemarketers selling insurance. At first, the department makes calls at random. Over time, the department identifies that young women in particular are inclined to buy insurance during the call. The department therefore starts calling only younger women, but this causes sales figures to drop. By conducting interviews, it is found that almost all of the women who buy the insurance would have bought it regardless of the phone call. On the other hand, men, for example, are less likely to buy the insurance -- but may be convinced through a phone call. The company therefore changes its strategy to never call young women, resulting in an increase in sales.}
\end{minipage}
\end{center}
The example shows that interventions should not necessarily target those with a high likelihood of buying (high risk) or even a high likelihood of buying given they receive the call (high risk given the intervention), but perhaps rather those for which the likelihood of buying increases the most by receiving a call (high impact given the intervention).

A systematic approach to impact evaluation, encompassing the three aforementioned questions, is causal inference -- because the questions are inherently causal: They ask what the impact of an intervention is on the claim{s} frequency and the claim{s} sizes, and how this effect varies depending on the covariates, henceforth denoted $W$.  In general, the causal effect of a prevention initiative (or treatment), henceforth denoted $A$, on some function of the future cash flow (or outcome), henceforth denoted $Y$, can only be inferred if there exist similar individuals and only some receive the treatment, while the others do not. Situations with detailed knowledge of the causal relationship between $Y$, $A$, and $W$ constitute an exception; here graphical models, for example, may provide other ways to infer the causal effect. Insurance applications, on the other hand, typically fall outside this exception due to the complexity of the underlying risks (life trajectories), the low signal-to-noise ratio, and restrictions on the availability of relevant covariates to the insurer. For an introduction to causal inference, we refer to the modern classics~\citet{Pearl:2009} and~\citet{Hernan:Robins:2020}. For applications to disability insurance, see Subsection~1.1.3 in~\citet{Sandqvist:2025}.

One situation where causal inference, however, remains possible is when the assumptions of \textit{no unmeasured confounding} and \textit{positivity} are satisfied. Assume for simplicity that $A$ is binary and that the outcome $Y$ is fully observed and not for instance subject to right-censoring. Further, let $Y^{(1)}$ and $Y^{(0)}$ correspond to what the outcome would have been if $A$ was set to $1$ and $0$, respectively, which are the so-called potential outcomes. Then $Y = Y^{(A)}$, and no unmeasured confounding states that
\begin{align*}
(Y^{(1)},Y^{(0)}) \indep A \mid W,
\end{align*}
so that treatment assignment and how the subject would react to treatment are independent conditional on $W$. Positivity states that $\mathbb{P}(A=1 \mid W=w)$ is uniformly bounded away from both one and zero as a function of $w$. In other words, for each possible realization of the covariates, there is a strictly positive probability both receiving and not receiving treatment. Under these assumptions, the conditions of a randomized controlled trial are satisfied for each possible realization of the covariates. Therefore, the conditional average treatment effect (CATE) becomes
\begin{align*}
\text{CATE}
&:=
\mathbb{E}[Y^{(1)} \mid W] - \mathbb{E}[Y^{(0)} \mid W] \\
&\hphantom{:}=
\mathbb{E}[Y \mid W, A=1]-\mathbb{E}[Y \mid W, A=0].
\end{align*}
{It can be difficult to account for unmeasured confounding related to the treatment itself, since individuals may opt out. In practice, it will therefore also be necessary to limit the analysis to measuring only the intention-to-treat effect~\citep{McCoy:2017}.}

Another situation in which causal inference is possible is if the condition for a \textit{regression discontinuity design} is satisfied, namely that there is a discontinuity in $w \mapsto \mathbb{P}(A=1 \mid W=w)$, where $W$ is now assumed to be one-dimensional. This creates a local randomized trial for subjects with covariates close to the discontinuity. If $W$ is one-dimensional and the discontinuity is at $w_0$, then the conditional average treatment effect at $w_0$ is identified as
\begin{align*}
\text{CATE}(w_0)
&:=
\mathbb{E}[Y^{(1)} \mid W=w_0] - \mathbb{E}[Y^{(0)} \mid W=w_0] \\
&\hphantom{:}=
\frac{\lim_{w \downarrow w_0}\mathbb{E}[Y \mid W=w] - \lim_{w \uparrow w_0}\mathbb{E}[Y \mid W=w]}{\lim_{w \downarrow w_0}\mathbb{P}(A=1 \mid W=w) - \lim_{w \uparrow w_0}\mathbb{P}(A=1 \mid W=w)}.
\end{align*}
An efficient estimation approach that applies to both situations with no unmeasured confounding and to regression discontinuity designs, and which allows the outcome to be affected by right-censoring, is proposed in~\citet{Sandqvist:2024}; see also Subsection~1.2.3 in~\citet{Sandqvist:2025}.

It becomes apparent from the above discussion that the data-generating mechanism has to admit a particular structure in order for causal inference to be possible. This may happen as a coincidental consequence of internal practices at the insurer. For example, if only subjects with an estimated risk over a certain threshold are targeted by a prevention initiative, one is perhaps in the setting of a regression discontinuity design. Similarly, if the assignment of treatment exclusively depends on a combination of covariates already available to the insurer and exogenous randomness, say how many claims were reported in the past month,  no unmeasured confounding and positivity may hold, and causal inference is again possible.

However, rather than just relying on happy accidents and pure luck, we advise actuaries to be actively involved in the design of prevention initiatives and in such a way that their impact becomes quantifiable and hence optimizable. Actuaries have a unique opportunity in this regard due to their close connection to the management of their organization, their knowledge of the insurance coverages' properties and functionalities, and their statistical expertise not least for data subject to sampling effects.

\section{  Data availability} \label{sec:data}
{

We have presented and discussed rather sophisticated modeling approaches, including multistate models that account for claim settlement as well as causal inference frameworks. However, the successful implementation of these methods relies on the availability and quality of data. In this section, we discuss the practical challenges associated with obtaining such data for disability insurance, contrasting the situation with the relatively data-rich environment for mortality risk.

Industry-wide data is, unfortunately, not as readily available for disability as it is for mortality. The reason for this is multifaceted. First, disability data constitutes sensitive personal information. In the European Union, the General Data Protection Regulation (GDPR) imposes strict limitations on how such data is stored, processed, and shared. Unlike mortality, which is typically a matter of public record, detailed health and employment data is protected, making the pooling of data across insurers for the construction of industry-wide tables significantly more difficult. Second, disability data lacks the standardization inherent to mortality data. While death is a binary and unambiguous event, disability is defined by complex insurance product criteria, which may vary between insurance providers and change over time. Consequently, data from one portfolio is not necessarily comparable to another, nor is historical data necessarily comparable to present data. This heterogeneity poses a significant hurdle for the amassing of data necessary to train and validate complex models. For these reasons, it is imperative that insurance providers are diligent about recording all relevant data, diligent about ensuring it is of high quality, and mindful of the nature and frequency of product changes.  

The individual-level models discussed in Section~\ref{sec:Modeling} require substantial higher data quantity and quality compared to traditional aggregate models. To calibrate models with advanced dependencies and trends, it is beneficial to have long, stationary time series. However, the disability risk landscape is rarely stationary. As noted in our discussion of trend risk, societal trends -- such as the surge in stress-related diagnoses, changes in social security legislation, and alterations in product design -- often render older records obsolete.  While individual models require larger samples and more granular data, they typically possess inherent explainability features and may therefore more easily be used to identify and accommodate distributional shifts; sometimes, they can even be made robust to such shifts.

Regarding the data required for causal inference, actuaries must be cautious when attempting to bridge the aforementioned data gaps with external medical or governmental studies. While clinical trials and public health studies offer valuable insights into disease progression and prevention, their results rarely generalize directly to the insurance setting. Insurance portfolios are subject to specific selection effects -- insured lives often exhibit different health characteristics than the general population -- and the insurance provider operates under a different authorization than medical professionals. Furthermore, the difference between clinical recovery and recovery of earning capacity may hinder the transferability of results.

In summary, the data infrastructure and governance required to support disability risk assessment remains a critical bottleneck. Overcoming these challenges requires not only advanced statistical techniques to handle small or biased datasets, but also necessitates a concerted effort to improve data standardization and collection practices within the constraints of privacy regulations. This is well within the scope of the actuarial profession, of which much must be expected in this area.}

\section{Outlook}  \label{sec:Conclusion}

Our review confirms that disability insurance is an area of high practical complexity, not least in Denmark, but also with great potential for meaningful insurance mathematical innovation. To stimulate research and promote better practices, we have sought to illustrate and systematize these complexities; to review and assess the adequacy of current approaches; and to identify avenues for future research and development. We find that modeling the interplay with public benefits and the associated implications for product design constitute an underdeveloped and promising topic for future research. Additionally, there is a need for empirical studies related to the design and impact evaluation of prevention initiatives. Actuaries must continue to play a crucial role in this context, embracing another dimension to their role, while maintaining strong ethics and safeguarding fairness. We aptly recall the message of~\citet{Barabas.etal:2018}: The core ethical debate surrounding risk assessments is not simply one of bias or accuracy, but one of purpose: away from prediction and towards prevention.

\section*{Conflicts of interest}

The expressed opinions are attributable solely to us and do not necessarily reflect the views of our past, current, and future employers.

\section*{Acknowledgements}

An earlier version of this paper appears as Chapter~2 in the PhD thesis of the second author, that is \citep[][Chapter 2]{Sandqvist:2025}. Oliver Lunding Sandqvist gratefully acknowledges support from the Innovation Fund Denmark under File No.~1044-00144B. Christian Furrer gratefully acknowledges support from Fynske Købstæders Fond. The authors would like to thank Mogens Steffensen {and three anonymous referees} for helpful comments on {earlier versions} of the {paper}.

\end{document}